\documentclass[aps,prl,showpacs,superscriptaddress,amssymb,twocolumn]{revtex4-1}
\usepackage{graphicx}
\usepackage{appendix}
\usepackage{color}
\usepackage[position=top,singlelinecheck=off,justification=raggedright,caption=false]{subfig}
\usepackage{epstopdf} 
\DeclareGraphicsExtensions{.pdf,.eps,.png,.jpg,.mps}
\usepackage[colorlinks=true,citecolor=blue,urlcolor=blue,linkcolor=blue]{hyperref}
\usepackage{bm,amsmath}
\usepackage{dcolumn}
\input{Qcircuit}

\def\BEq{\begin{equation}}
\def\EEq{\end{equation}}
\def\BEqA{\begin{eqnarray}}
\def\EEqA{\end{eqnarray}}
\def\BW{\begin{widetext}}
\def\EW{\end{widetext}}

\begin{document}

\title{A leakage-resilient approach to fault-tolerant quantum computing with superconducting elements}

\author{Joydip Ghosh}
\email{ghoshj@ucalgary.ca}
\affiliation{Institute for Quantum Science and Technology, University of Calgary, Calgary, Alberta T2N 1N4, Canada}
\author{Austin G. Fowler}
\email{austingfowler@gmail.com}
\affiliation{Department of Physics, University of California, Santa Barbara, California 93106, USA}
\affiliation{Centre for Quantum Computation and Communication Technology, School of Physics, The University of Melbourne, Victoria 3010, Australia}

\date{\today}

\begin{abstract}

Superconducting qubits, while promising for scalability and long coherence times, contain more than two energy levels, and therefore are susceptible to errors generated by the \emph{leakage} of population outside of the computational subspace. Such leakage errors constitute a prominent roadblock towards Fault-Tolerant Quantum Computing (FTQC) with superconducting qubits. FTQC using topological codes is based on sequential measurements of multi-qubit stabilizer operators. Here, we first propose a leakage-resilient procedure to perform repetitive measurements of multi-qubit stabilizer operators, and then use this scheme as an ingredient to develop a leakage-resilient approach for surface code quantum error correction with superconducting circuits. Our protocol is based on SWAP operations between data and ancilla qubits at the end of every cycle, requiring read-out and reset operations on every physical qubit in the system, and thereby preventing persistent leakage errors from occurring.

\end{abstract}

\pacs{03.67.Lx, 03.67.Pp, 85.25.-j}    

\maketitle

Recent years have witnessed remarkable progress in quantum computing with superconducting components, as far as the scalability and coherence times are concerned \cite*{Barends2014,PhysRevLett.109.060501,2013arXiv1311.6330C}. With major advances in designing scalable qubits \cite*{PhysRevLett.111.080502,Barends2014,2014arXiv1402.7367C} and high-fidelity quantum gates \cite*{Barends2014,Motzoi2009,PhysRevA.87.022309,2014arXiv1402.5467M,0953-2048-27-1-014001}, a significant fraction of this research endeavor is now directed towards Fault-Tolerant Quantum Computing (FTQC) with superconducting devices \cite*{PhysRevA.86.062318,2013arXiv1311.6330C}. FTQC via topological error-correcting codes, such as the surface code \cite*{dennis:4452}, requires sequential measurements of multi-qubit stabilizer operators \cite*{PhysRevA.86.032324}. The fluctuations in the measurement outcomes of such stabilizer operators generate characteristic signatures for various discrete Pauli errors, thereby rendering the error-correcting scheme robust against error models described by a Pauli channel \cite*{PhysRevA.86.032324}.

Superconducting qubits comprise more than two energy levels that are often utilized to design two-qubit entangling gates, such as a controlled-$\sigma^{z}$ (CZ) gate \cite*{PhysRevLett.91.167005,PhysRevA.87.022309,0953-2048-27-1-014001,2014arXiv1402.5467M}. Apart from the decoherence, therefore, superconducting qubits also suffer from errors due to \emph{leakage} of population outside of the computational subspace, often referred to as leakage errors \cite{Aliferis2007,Fong2011,PhysRevA.88.042308,PhysRevA.88.062329,2014arXiv1410.8562S}.
While decoherence-induced errors can be approximated by a Pauli channel \cite*{Geller&ZhouPRA13}, leakage errors lack such a description, thereby compromising the fault-tolerance offered by the standard stabilizer-based schemes, unless supplemented by a Leakage Reduction Unit (LRU) \cite{Aliferis2007,2014arXiv1410.8562S}. 

It has been shown recently with numerical simulations, how persistent leakage errors in superconducting circuits destroy an ancilla-assisted qubit-measurement scheme producing random fluctuations in the output of the ancilla qubit \cite*{PhysRevA.88.062329}.
In this Rapid Communication, we first propose a scheme for multi-qubit stabilizer-measurement, which is resilient to such leakage errors, and then develop a scalable leakage-resilient protocol for surface code quantum error correction.

\begin{figure}[htb]
\centering
\subfloat[]{
\label{fig:stdScheme}
\includegraphics[angle=0,width=\linewidth]{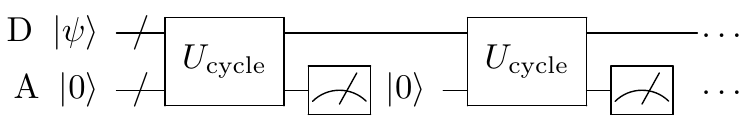}
}
\\
\subfloat[]{
\label{fig:lrScheme}
\includegraphics[angle=0,width=\linewidth]{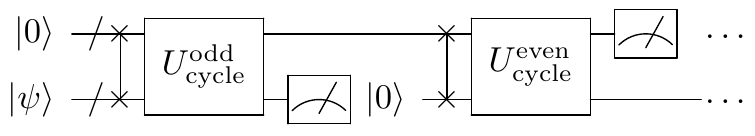}
}
\caption{Repetitive measurements of stabilizer operators via (a) standard scheme, and (b) our SWAP-based scheme. The ancilla register (A) gets measured and initialized at each cycle in the standard scheme, while the data register (D) never gets reset. Both the registers are measured in our scheme at alternate cycles. $U_{\rm cycle}$ denotes the sequence of gate operations required for the stabilizer measurement in the standard scheme, and those for SWAP-based scheme are denoted by $U^{\rm odd}_{\rm cycle}$ and $U^{\rm even}_{\rm cycle}$ for odd and even cycles repsectively.}
\label{fig:generalScheme}
\end{figure}

A schematic diagram of the standard and our SWAP-based approach for repetitive measurements of stabilizer operators is shown in Fig.~\ref{fig:generalScheme}. In the standard approach, the data qubit register never gets measured, and therefore, any leaked qubit in the data register remains leaked for many cycles until it undergoes relaxation due to decoherence or leaks back to the computational subspace. As shown in Ref.~\cite*{PhysRevA.88.062329}, such leaked data qubits generate random noises in the measurement outcomes of ancilla qubits, effectively spoiling the entire scheme. Note that, even though it is possible to detect the location of a leakage error from some sequential measurements of the ancilla qubits, it is not possible to correct it applying a single-qutrit unitary operation on the leaked data qubit, as the exact state of the data qubit remains unknown. In order to circumvent the harmful consequences of leakage errors, we must, therefore, resort to some projective measurement on the data qubits as well, which is absent in the standard scheme, as shown in Fig.~\ref{fig:stdScheme}.

In our leakage-resilient protocol, we supplement the standard approach with SWAP operations at the end of every measurement-cycle, as shown in Fig.~\ref{fig:lrScheme}. The SWAP gates exchange the roles of the data-register and the ancilla-register, requiring us to measure and initialize every qubit in alternate cycles, thereby eliminating the possibility of persistent leakage errors without compromising the stabilizer-measurement scheme.

For both the standard and our SWAP-based approach, we assume that the readout of ancilla qubits can resolve $\ket{0}$, $\ket{1}$ and $\ket{2}$ states separately \cite*{Neeley07082009}. This assumption, however, is not a requirement for our scheme to work, as persistent leakage errors on the data qubits are completely independent of a leakage error on an ancilla qubit, and a leaked ancilla qubit always gets reset right after the readout in any case. Since the outcomes of the ancilla qubits are used for predicting the state encoded in the data-register, a leakage error on an ancilla qubit in a given cycle only amounts to an erroneous prediction for that particular cycle only. It is, therefore, equivalent for our purpose if we resolve $\ket{2}$ states in the ancilla-readout or simply map it to a specific computational state, such as the $\ket{1}$ state, which needs to be done anyway for predicting the data-register-state, as discussed later.

\begin{figure}[htb]
\centering
\subfloat[]{
\includegraphics[angle=0,width=0.33\linewidth]{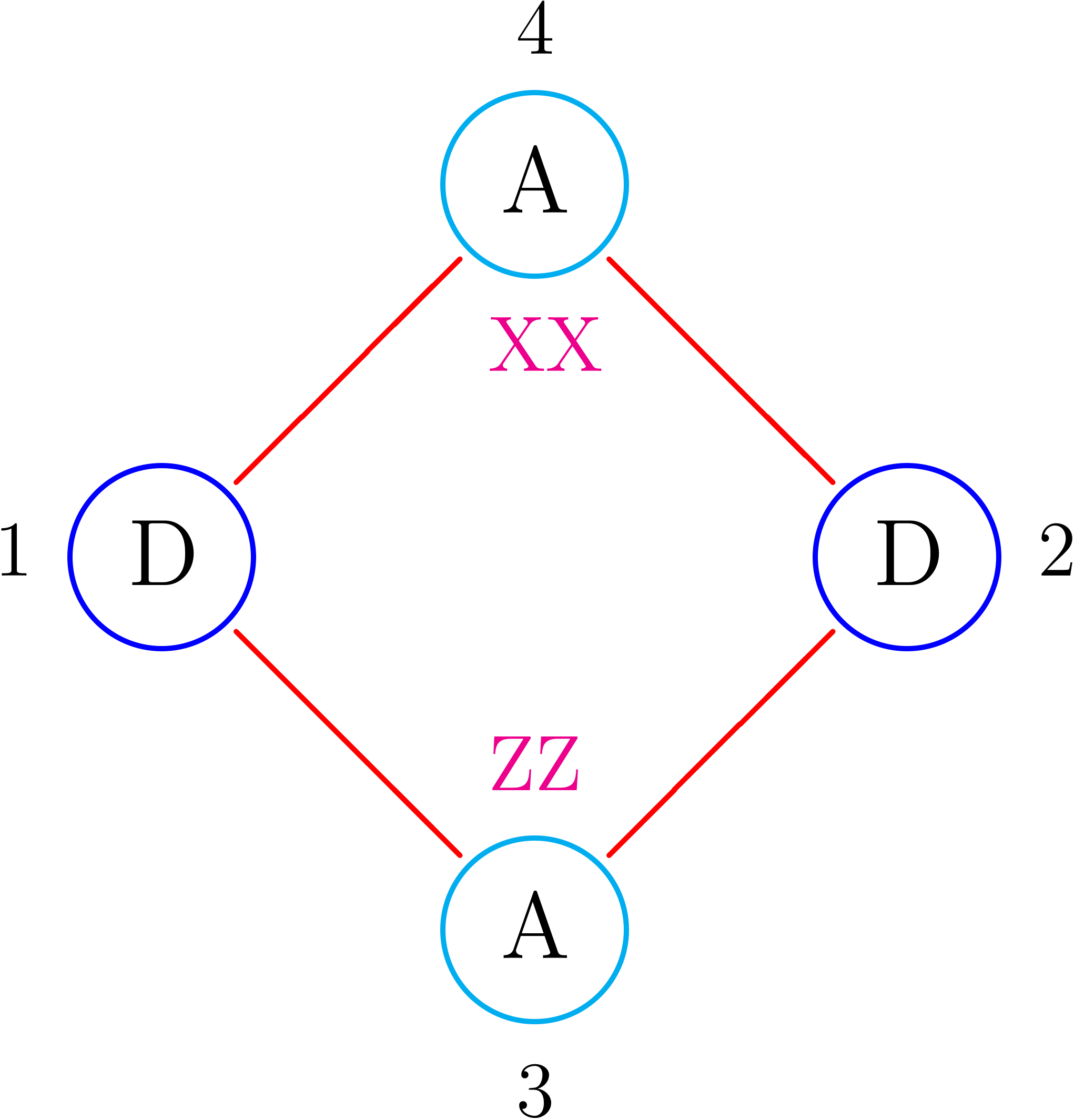}
\label{fig:architecture}
}
\subfloat[]{
\includegraphics[angle=0,width=0.65\linewidth]{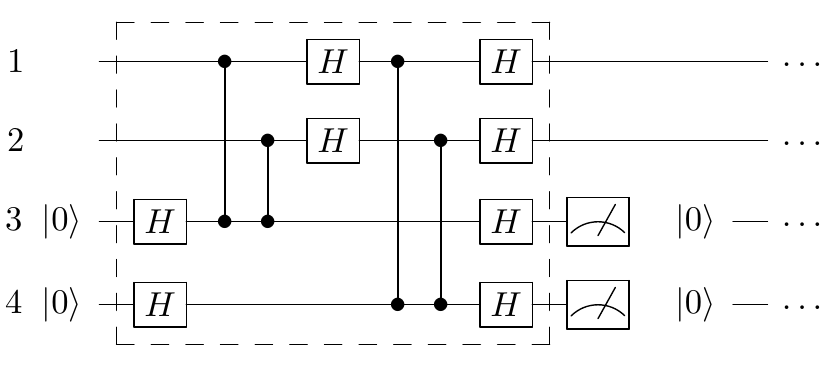}
\label{fig:standardCircuit}
}
\caption{(a) (Color online) A schematic diagram of the architecture for measuring two-qubit stabilizer operators. The circles denote superconducting qubits (data qubits denoted by `D' and ancilla qubits denoted by `A'), and lines denote required nearest-neighbor couplings. The stabilizer operators, $ZZ$ and $XX$, are measured via qubit-3 and qubit-4 respectively. (b) Standard scheme for repetitive measurements of two-qubit stabilizer operators $XX$ and $ZZ$. $H$ denotes the Hadamard gate, the vertical lines connected by filled circles denote CZ gates, and the numbers denote the indices for the qubits in the same order as shown in Fig.~\ref{fig:architecture}. The gates inside the dashed rectangle represent $U_{\rm cycle}$ in Fig.~\ref{fig:stdScheme}.}
\end{figure}

Here, we illustrate the advantage of our measurement protocol assuming a model where both the data- and ancilla-registers consist of two superconducting qubits, and we measure two 2-qubit stabilizer operators, $XX$ and $ZZ$, repetitively for many cycles. Fig.~\ref{fig:architecture} shows a schematic diagram of our architecture. The standard scheme for stabilizer measurement for this model is shown in Fig.~\ref{fig:standardCircuit} \cite*{Geller&ZhouPRA13}. Note that, if we encode any of the four Bell states in the data register, then under ideal gate operations, the final states before reinitialization are given by,

\BEq
\label{eq:BellStates}
\begin{array}{l}
\left(\frac{\ket{00}+\ket{11}}{\sqrt{2}}\right)\otimes \ket{00} \longmapsto \left(\frac{\ket{00}+\ket{11}}{\sqrt{2}}\right)\otimes \ket{00}, \\
\left(\frac{\ket{00}-\ket{11}}{\sqrt{2}}\right)\otimes \ket{00} \longmapsto \left(\frac{\ket{00}-\ket{11}}{\sqrt{2}}\right)\otimes \ket{01}, \\
\left(\frac{\ket{01}+\ket{10}}{\sqrt{2}}\right)\otimes \ket{00} \longmapsto \left(\frac{\ket{01}+\ket{10}}{\sqrt{2}}\right)\otimes \ket{10}, \\
\left(\frac{\ket{01}-\ket{10}}{\sqrt{2}}\right)\otimes \ket{00} \longmapsto \left(\frac{\ket{01}-\ket{10}}{\sqrt{2}}\right)\otimes \ket{11},
\end{array}
\EEq
which essentially means that in absence of any error under circuit~\ref{fig:standardCircuit}, the four different Bell states in the data register are stabilized by the operators $XX$ and $ZZ$ as their simultaneous eigenstates corresponding to the four different possible combinations. Without any loss of generality, in this work we assume the Bell state $(\ket{00}+\ket{11})/\sqrt{2}$ as our encoded initial state in the data register and compare our protocol against the standard scheme simulating the circuits numerically.

\begin{figure}[htb]
\centering
\includegraphics[angle=0,width=\linewidth]{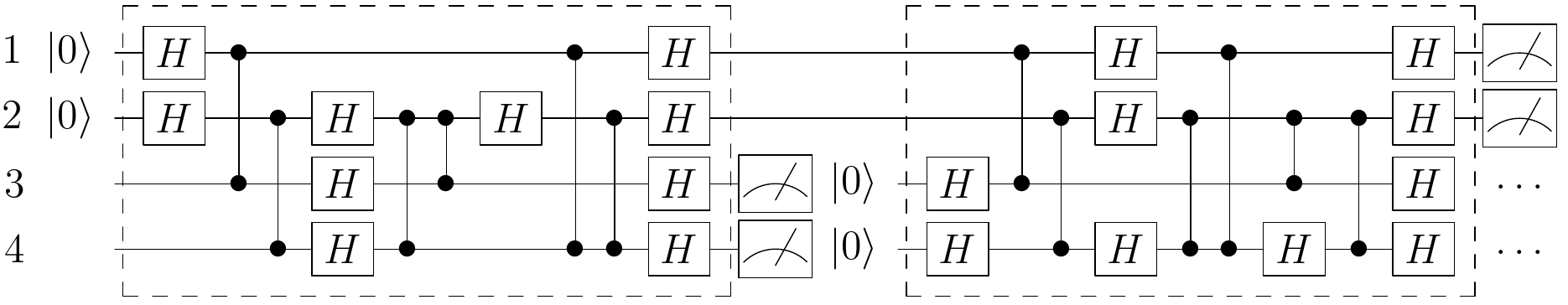}
\caption{The leakage-resilient scheme for stabilizer measurement where SWAP gates are replaced by CZ and Hadamard gates as shown in Eq.(\ref{eq:SWAPCZ}). The gates inside the left dashed rectangle are repeated for odd cycles, while the gates inside the right rectangle are repeated for even cycles.}
\label{fig:reducedSwapScheme}
\end{figure}

In order to render this repetitive stabilizer measurement scheme leakage-resilient, we introduce SWAP operations between the data and ancilla registers as shown in Fig.~\ref{fig:lrScheme}. SWAP operations between two quantum registers essentially mean sequential SWAP gates between the $k^{\rm th}$ qubits in both the registers, for all $k$. In order to not transfer or propagate the leakage errors across the circuit via the superconducting qubits, we express SWAP operations (between $\ket{0}$ and an arbitrary state $\ket{\psi}$) as,
\begin{equation}
\small
\centering
\begin{array}{l}
\mbox{
\Qcircuit @C=0.7em @R=0.3em {
\lstick{\ket{\psi}} & \qswap        & \qw  &     &   & \qw     & \ctrl{2} & \gate{H} & \ctrl{2} & \gate{H} & \qw \\
                       &  \qwx           &        & \equiv  &   &               &            &                &            &               &        \\
\lstick{\ket{0}} & \qswap\qwx & \qw &      &   & \gate{H} & \ctrl{0} & \gate{H} & \ctrl{0} & \qw         & \qw
}
}
\end{array}
\label{eq:SWAPCZ}
\end{equation}
We note that the substitution (\ref{eq:SWAPCZ}), in fact, introduces a negligible overhead in circuit depth because of internal cancellations, and the reduced circuit for our SWAP-based stabilizer measurement scheme is shown in Fig.~\ref{fig:reducedSwapScheme}.

In order to estimate the dominant contribution for the leakage errors, we model the superconducting qubits as three-level systems (or qutrits) and parametrize the non-ideal single- and two-qutrit quantum gates, as discussed in Ref.~\cite*{PhysRevA.88.062329}. An ideal Hadamard gate for a qutrit (in the basis $\{\ket{0}, \ket{1}, \ket{2}\}$) is given by,
\begin{equation}
H = \left[\begin{array}{ccc}
 \frac{1}{\sqrt{2}} &  \frac{1}{\sqrt{2}} & 0\\
 \frac{1}{\sqrt{2}} &  -\frac{1}{\sqrt{2}} & 0 \\
 0 & 0 & 1 \\
\end{array}\right],
\label{Hadamard}
\end{equation}
which is equivalent to the standard single-qubit Hadamard gate acting on the computational subspace, and an Identity acting on the $\ket{2}$ state. 
The leakage errors produced by the non-ideal single-qubit gates are \emph{local}, leaving no visible signature in the outcomes of ancilla measurement, and therefore, remain undetectable under the standard scheme. In our SWAP-based approach, as all qubits are repeatedly measured and initialized, such local leakage errors get automatically erased anyway. 
However, since single-qubit gates can be done with fidelities an order of magnitude higher than that of two-qubit gates \cite*{Motzoi2009,Kell14}, we argue that it is sufficient to consider the leakage errors generated by the two-qubit gates alone for the purpose of this work.

The two-qubit CZ gate considered here is performed by tuning and detuning the qubit-frequencies so as to mix the population in the avoided level-crossing of $\ket{11}$ and $\ket{20}$ eigenstates, such that the $\ket{11}$ state acquires a phase of angle $\pi$ \cite*{Barends2014,PhysRevLett.91.167005,PhysRevA.87.022309,2014arXiv1402.5467M}. The dominant error for an avoided-crossing-based CZ gate is generated by the residual non-adiabatic population exchanges in single-excitation ($\{\ket{01},\ket{10}\}$) and double-excitation ($\{\ket{02},\ket{20},\ket{11}\}$) subspaces. As shown in Ref.~\cite*{PhysRevA.88.062329}, such a CZ gate can be parametrized with two generators, $S$ (generates the ideal part) and $S'$ (generates the first-order error terms), where the parameters are chosen from a full-scale Hamiltonian simulation. The generators $S$ and $S'$ can be thought of as Hermitian matrices generating the unitary non-ideal CZ gate, $U_{\rm CZ}=e^{i\left(S+S'\right)}$.
In the two-qutrit tensor-product basis
$\big\lbrace \ket{00},\ket{01},\ket{02},\ket{10},\ket{11},\ket{12},\ket{20},\ket{21},\ket{22}  \big\rbrace$,
the generator $S$ is given by \cite*{PhysRevA.88.062329},
\BEq
\label{eq:SCZ}
S=\text{diag}(0,0,\xi_{1},0,\pi,\xi_{2},\pi,\xi_{3},\xi_{4}),
\EEq
and the generator $S'$ can be represented in single-, double-, and triple-excitation subspaces as \cite*{PhysRevA.88.062329},
 \BEq
\label{eq:SprimeCZ}
\begin{array}{l}
S'_{\{\ket{01},\ket{10}\}}=\left[\begin{array}{cc}
 \zeta_1 & i\chi_{1}e^{i\phi_1} \\
 -i\chi_{1}e^{-i\phi_1} & \zeta_2 \\
 \end{array}\right], \\
 
 S'_{\{\ket{02},\ket{11},\ket{20}\}}=\left[\begin{array}{ccc}
 0 & i\chi_{2}e^{i\phi_2} & 0 \\
 -i\chi_{2}e^{-i\phi_2} & \zeta_3 &  i\chi_{3}e^{i\phi_3} \\
 0 & -i\chi_{3}e^{-i\phi_3} & \zeta_4 \\
 \end{array}\right], \\
 
 S'_{\{\ket{12},\ket{21}\}}=\left[\begin{array}{cc}
 0 & i\chi_{4}e^{i\phi_4} \\
 -i\chi_{4}e^{-i\phi_4} & 0 \\
 \end{array}\right],
\end{array}
\EEq
where the subscripts in (\ref{eq:SprimeCZ}) denote the basis sets for the corresponding subspaces. 
 
Since the CZ gate is performed by mixing the population in the avoided level-crossing between $\ket{11}$ and $\ket{20}$ states, we have $\bra{11}S\ket{11}=\bra{20}S\ket{20}=\pi$ in Eq.(\ref{eq:SCZ}). The remaining parameters $\xi_{1-4}$ are the dynamical phases acquired by the corresponding basis states and are assumed to be random numbers between $0$ and $2\pi$ for our simulation. The parameters in $S'$, $\chi_{1-4}$ and $\zeta_{1-4}$, are assumed to be small ($\sim 10^{-2}$), while the angles $\phi_{1-4}$ take arbitrary values between $0$ and $2\pi$. The diagonal elements of $S'$ corresponding to $\ket{02}$, $\ket{12}$, $\ket{21}$, and $\ket{22}$ states are `0', as the residual phases across these states are already absorbed in the definition of dynamical phases $\xi_{1-4}$. The parameters $\chi_{1}$ and $\chi_{4}$ cause population-transfer in the single- and triple-excitation subspaces respectively, while $\chi_{2,3}$ are responsible for mixing of population in the double-excitation subspace. Since the population transfer probabilities scale with $|\chi_{i}|^{2}$ \cite*{PhysRevA.88.062329}, our choice of parameters imply gate errors $\sim 10^{-4}$, which is consistent with what has been obtained from a full-scale simulation of the control-Hamiltonian for current gate-design schemes \cite*{PhysRevA.87.022309,2014arXiv1402.5467M}.

Having parametrized the required quantum gates, we now simulate the quantum circuits shown in Fig.~\ref{fig:standardCircuit} and in Fig.~\ref{fig:reducedSwapScheme}. In order to compare our scheme against the standard approach, we here assume the no-decoherence limit (i.e., $T_{\rm 1,2}\rightarrow \infty$). We emphasize that the introduction of decoherence in our calculation only amounts to some more randomly occurred `steps' in the readout values that are neither relevant for, nor influence the conclusions of this work.

\begin{figure}[htb]
\centering
\includegraphics[angle=0,width=\linewidth]{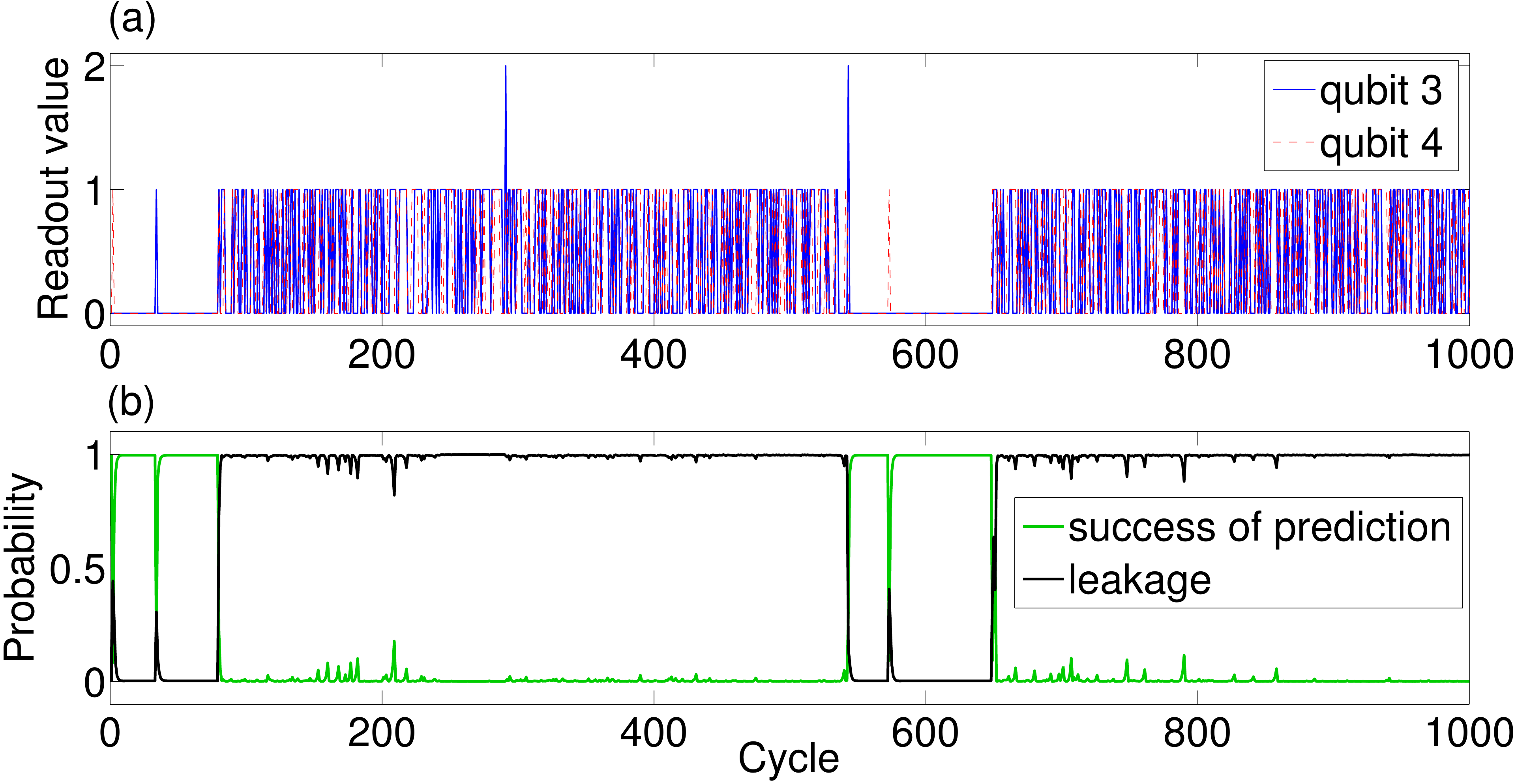}
\caption{(Color online) Results for the standard approach to repetitive measurements of two-qubit stabilizer operators, $XX$ (dashed red) and $ZZ$ (solid blue). (a) The readouts of two ancilla qubits (qubits 3 and 4) are shown for various consecutive cycles. (b) The solid black curve shows the probability that either of the two data qubits is in the $\ket{2}$ state at the end of every measurement cycle. The green (solid gray) curve shows the overlap between the state encoded in the data register and the prediction of it from the corresponding ancilla outputs at the end of each cycle.}
\label{fig:StandardData}
\end{figure}

\begin{figure}[htb]
\centering
\includegraphics[angle=0,width=1\linewidth]{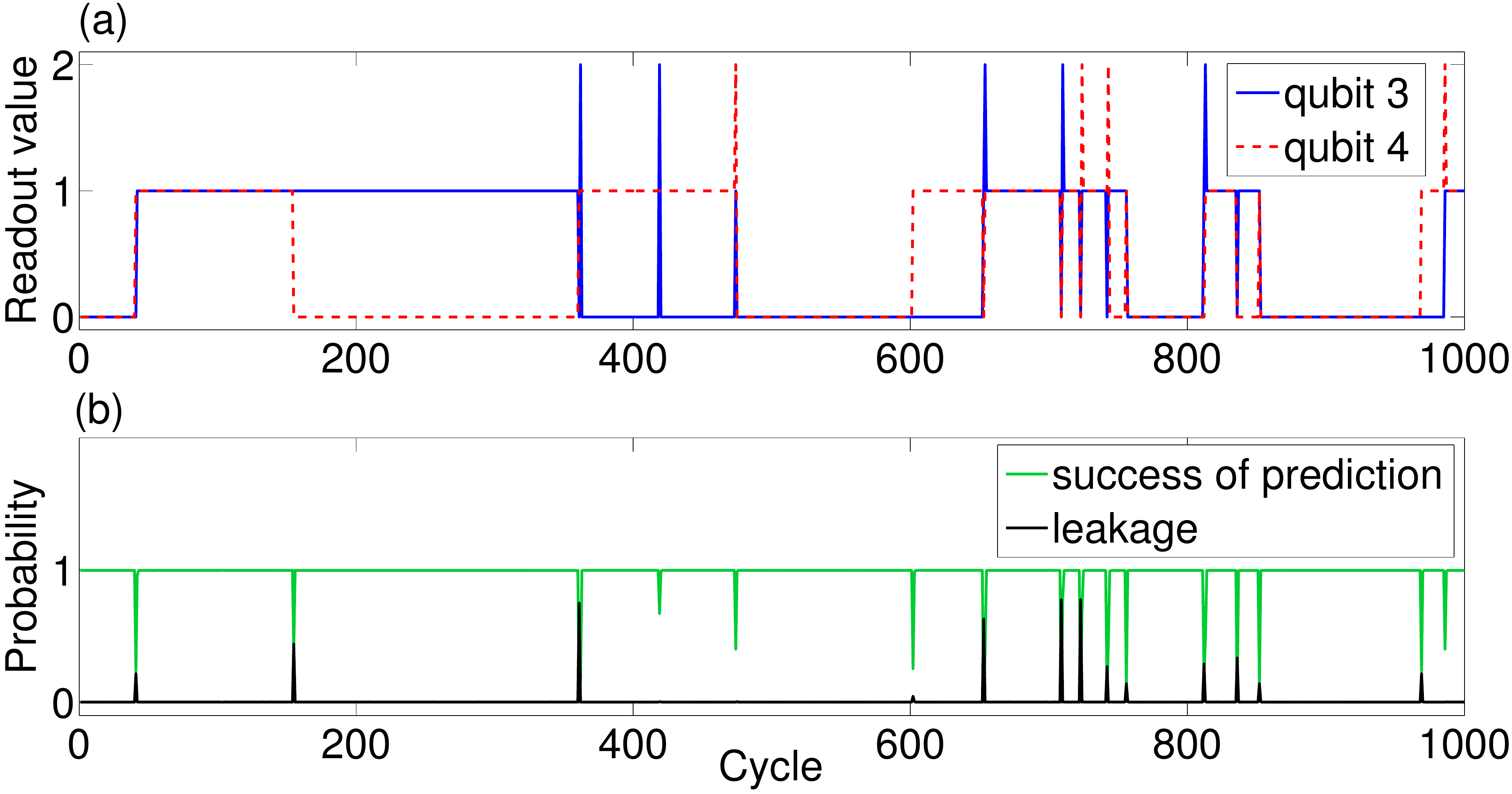}
\caption{(Color online) Results for our SWAP-based approach to repetitive measurements of two-qubit stabilizer operators, $XX$ (dashed red) and $ZZ$ (solid blue). (a) The readouts of two ancilla qubits (qubits 3 and 4) are shown for various consecutive cycles. (b) The solid black curve shows the probability that either of two data qubits is in the $\ket{2}$ state at the end of every measurement cycle. The green (solid gray) curve shows the overlap between the state encoded in the data register and the prediction of it from the corresponding ancilla outputs at the end of each cycle.}
\label{fig:SwapData}
\end{figure}

The simulation of the standard approach (Fig.~\ref{fig:standardCircuit}) is shown in Fig.~\ref{fig:StandardData}. The measurement outcomes from the ancilla qubits (qubits 3 and 4) for many consecutive cycles are shown in Fig.~\ref{fig:StandardData}a. In the presence of leakage errors, we observe regions having random and rapid fluctuations in the ancilla outcomes, a characteristic signature for a leakage error on either of the data qubits~\cite*{PhysRevA.88.062329}. In order to show the connection between this noise and the data-qubit leakage error more explicitly, in Fig.~\ref{fig:StandardData}b we plot the probability (black curve) that either of the data qubits is leaked at the end of every cycle. For cycles with rapid fluctuations in the outcomes of the ancilla qubits, we also observe a near-unit probability for leakage errors, clearly signifying the destructive consequences of data-qubit-leakage for the standard stabilizer-measurement scheme. Using Eq.(\ref{eq:BellStates}), it is possible to predict the quantum state of the data-qubit register, based on the outcomes obtained in the ancilla qubits. If the ancilla is in $\ket{2}$ state, we need to map it to some computational state for the purpose of such a prediction, and we assume $\ket{2} \mapsto \ket{1}$ in this work (while this choice is arbitrary, it does not change the conclusions of this work).  In Fig.~\ref{fig:StandardData}b, we also plot the probability of success for such predictions (green curve) at the end of each cycle. It is observed that this success-probability is enormously compromised for many consecutive cycles where the data qubits remain leaked, essentially indicating a catastrophic failure of the standard scheme under leakage errors.

We also simulate our SWAP-based scheme (Fig.~\ref{fig:reducedSwapScheme}) and the results are shown in Fig.~\ref{fig:SwapData}. Fig.~\ref{fig:SwapData}a shows the outputs of the ancilla qubits for many consecutive cycles and, unlike standard protocol, no rapid random fluctuations are observed in the ancilla outcomes for this case. The probability of a leakage error in the data register is shown (black curve) in Fig.~\ref{fig:SwapData}b. In contrast with the standard scheme, we only observe isolated peaks, which means even if there is a data-qubit leakage-event in one cycle, it gets completely removed in subsequent measurement cycles, as all qubits are measured in alternate cycles. Fig.~\ref{fig:SwapData}b also shows the probability of successful prediction (green curve) of the two-qubit state encoded in the data-qubit register. Notice that the predictions only get compromised whenever there is a leakage error either in the data-register or in the ancilla-register, and since the leakage errors are isolated, so are the failure probabilities. The discrete well-separated peaks in the leakage error plot explicitly signify the resilience of our SWAP-based scheme against leakage errors. It is possible to suppress the number density of such peaks even further with better optimization techniques \cite*{0953-2048-27-1-014001}, while even a single data-qubit leakage-event ruins the entire stabilizer measurement for the standard approach, as the leaked qubit remains leaked for a long time in that case.

\begin{figure}[htb]
\centering
\includegraphics[angle=0,width=\linewidth]{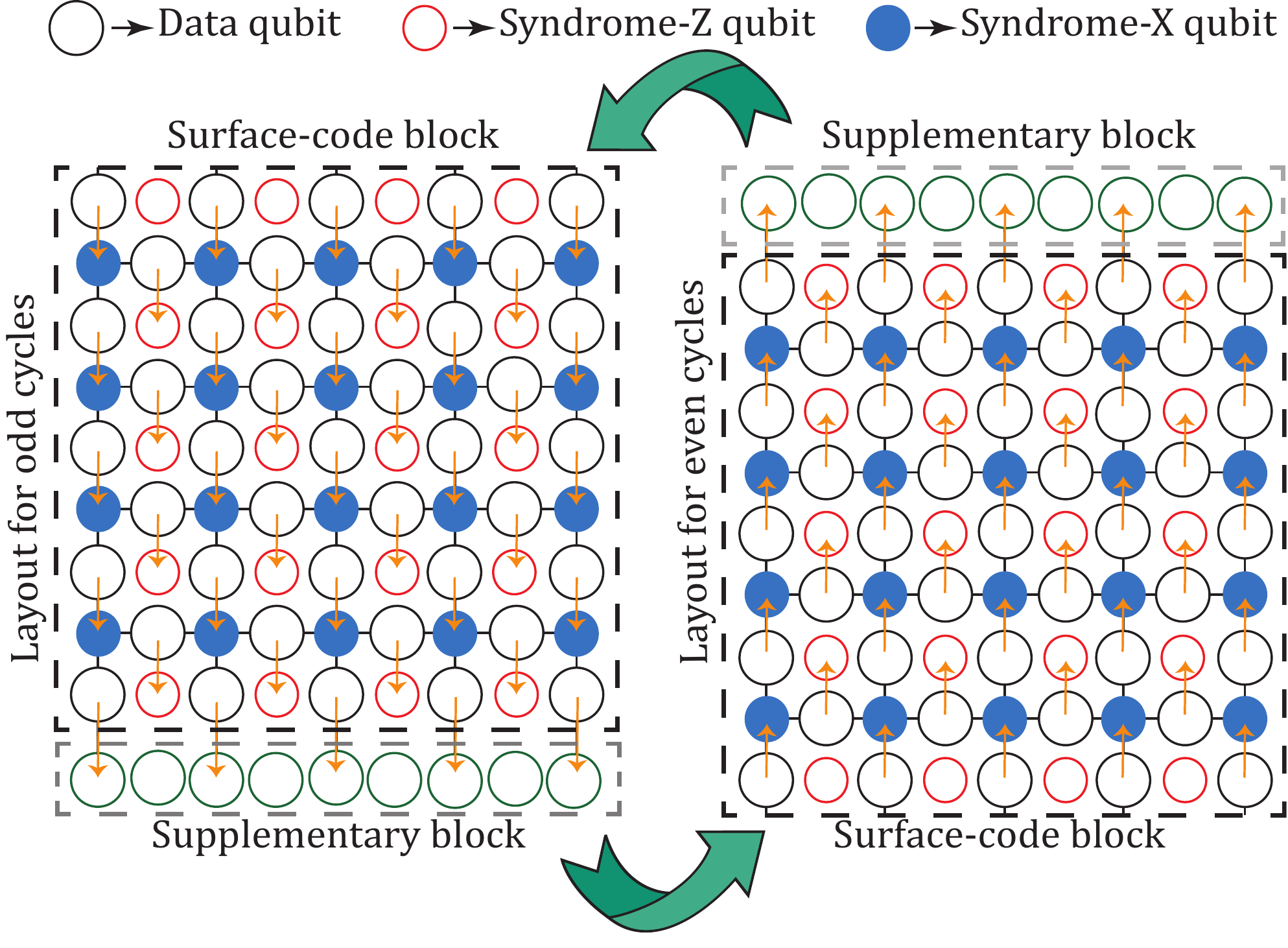}
\caption{(Color online) A schematic diagram of a leakage-resilient approach to surface code quantum error correction. The vertical arrows denote the directions for the transfer for quantum state from data qubits to syndrome qubits via SWAP gates after each cycle.}
\label{fig:SwapBasedSurfaceCode}
\end{figure}

Now, we outline how to devise a full-scale leakage-resilient scheme for FTQC using our SWAP-based stabilizer measurement protocol as an ingredient. Fig.~\ref{fig:SwapBasedSurfaceCode} shows a schematic diagram of this approach for distance-five surface code \cite*{PhysRevA.86.032324}, while the generalization of this scheme is trivial for any arbitrary distance. The black circles inside the black dashed rectangles represent the data qubits for a surface code architecture, while the red (blue) circles denote syndrome qubits that are used as ancilla for measuring three- or four-qubit Pauli Z (X) operators for the nearest-neighbor data-qubits. 
In order to implement the SWAP-based scheme, the quantum state encoded in the data qubits needs to be transferred to the syndrome qubits after every surface-code-cycle. However, the different arrangements of the data and syndrome qubits prevent us from completing this task, unless we use some additional physical qubits, here referred to as \emph{supplementary} qubits (shown inside the dashed gray rectangles). For a surface code with distance-$d$, we need $2d-1$ supplementary qubits arranged as an extra row, as shown in Fig.~\ref{fig:SwapBasedSurfaceCode}. After each odd (even) cycle, the quantum state encoded in the data qubits gets transferred to the syndrome qubits via SWAP operations as denoted by downward (upward) arrows, while the lowermost (uppermost) row of the surface-code block requires the supplementary qubits in order to complete the entire transfer process, and the uppermost (lowermost) row becomes the supplementary block for the next cycle. Note that all these SWAP gates can be performed simultaneously, and, therefore, the additional time-cost of our scheme is independent of the distance, while the number of required supplementary qubits scales only linearly with $d$. Repetition of these steps along with the standard surface-code-cycle requires readout and reset operations on all the physical qubits, thereby rendering the quantum computing protocol not only fault-tolerant, but also leakage-resilient.

In summary, we have devised a leakage-resilient protocol for repetitive measurements of multi-qubit stabilizer operators, and shown how to exploit this protocol to devise a leakage-resilient FTQC scheme with superconducting elements. In the standard approach for surface code error correction, the data qubits never get measured, and therefore a leaked data qubit remains leaked for many consecutive cycles, producing random fluctuations in the syndrome measurements, and thereby compromising the fault-tolerance. Our scheme relies on SWAP operations between the data and the syndrome qubits, requiring us to perform readout and reset operations on every physical qubit in the quantum circuit, which essentially eliminates the possibility of long-lived leakage errors. While our protocol is readily applicable to superconducting implementation of topological error correction performed via avoided-crossing-based two-qubit CZ gates, the idea of carrying out projective measurements on all the qubits at alternate cycles should be useful for many other models of leakage errors \cite{2014arXiv1410.8562S}. Computation of threshold for a surface or toric code quantum computing using our scheme will be considered as a possible future research direction.

We thank Michael Geller for helpful discussions. This research was funded by the US Office of the Director of National Intelligence (ODNI), Intelligence Advanced Research Projects Activity (IARPA), through the US Army Research Office grant No.~W911NF-10-1-0334. All statements of fact, opinion or conclusions contained herein are those of the authors and should not be construed as representing the official views or policies of IARPA, the ODNI, or the US Government. J.G. gratefully acknowledges the financial support from NSERC, AITF and University of Calgary's Eyes High Fellowship Program.

\bibliography{leakageTolerance}

\end{document}